\documentclass[conference]{IEEEtran}
\IEEEoverridecommandlockouts

\usepackage{cite}
\usepackage{amsmath,amssymb,amsfonts}
\usepackage{algorithmic}
\usepackage{graphicx}
\usepackage{textcomp}
\usepackage{xcolor}
\usepackage{pifont}
\usepackage{booktabs}
\usepackage{array}
\usepackage{tikz}
\usetikzlibrary{shapes.geometric, arrows}

\newcommand{\dpv}[1]{\mathbf{#1}}

\usepackage{multirow}
\def\BibTeX{{\rm B\kern-.05em{\sc i\kern-.025em b}\kern-.08em
    T\kern-.1667em\lower.7ex\hbox{E}\kern-.125emX}}
\begin{document}

\title{ASPO: Constraint-Aware Bayesian Optimization for FPGA-based Soft Processors}

\author{
Haoran Wu$^{*}$, Ce Guo$^{\dagger}$, Wayne Luk$^{\dagger}$, Robert Mullins$^{*}$ \\
$^{*}$Department of Computer Science and Technology, University of Cambridge, Cambridge, UK \\
$^{\dagger}$Department of Computing, Imperial College London, London, UK \\
Emails: hw691@cam.ac.uk, \{c.guo, w.luk\}@imperial.ac.uk, robert.mullins@cl.cam.ac.uk
}

% \author{Anonymous Submission}

\maketitle

\begin{abstract}
Bayesian Optimization (BO) has shown promise in tuning processor design parameters. However, standard BO does not support constraints involving categorical parameters such as types of branch predictors and division circuits. In addition, optimization time of BO grows with processor complexity, which becomes increasingly significant especially for FPGA-based soft processors. This paper introduces ASPO, an approach that leverages disjunctive form to enable BO to handle constraints involving categorical parameters. Unlike existing methods that directly apply standard BO, the proposed ASPO method, for the first time, customizes the mathematical mechanism of BO to address challenges faced by soft-processor designs on FPGAs. Specifically, ASPO supports categorical parameters using a novel customized BO covariance kernel. It also accelerates the design evaluation procedure by penalizing the BO acquisition function with potential evaluation time and by reusing FPGA synthesis checkpoints from previously evaluated configurations. ASPO targets three soft processors: RocketChip, BOOM, and EL2 VeeR. The approach is evaluated based on seven RISC-V benchmarks. Results show that ASPO can reduce execution time for the ``multiply'' benchmark on the BOOM processor by up to 35\% compared to the default configuration. Furthermore, it reduces design time for the BOOM processor by up to 74\% compared to Boomerang, a state-of-the-art hardware-oriented BO approach.
\end{abstract}


\section{Introduction}

Soft processors are instruction processors whose architecture and behavior are captured in software. They can be deployed on reconfigurable fabrics such as FPGAs~\cite{soft_proc_intro} and support different application programs without lengthy place-and-route. A soft processor is a parametric hardware design where the characteristics of microarchitectural components, such as the number of cache lines and the choice of the division circuit, are determined by a set of tunable configuration parameters defining the design space for the soft processor. The design process involves the selection of appropriate values for each parameter, enabling it to be optimized for specific tasks.

Manually optimizing soft processors for enhanced performance in specific tasks is impractical due to their vast and complex design space. This complexity arises from the high dimensionality of the parameter space and the non-linear relationships between parameters and performance. Additionally, many parameters are interdependent, requiring careful coordination to achieve optimal configurations. The challenge is further compounded by the lengthy evaluation process, which involves FPGA synthesis and significantly limits the number of configurations that can be tested during the tuning process.

To address these issues, we introduce ASPO, \textbf{A}utomated \textbf{S}oft \textbf{P}rocessors \textbf{O}ptimization, a soft processor parameter optimization platform based on Bayesian Optimization (BO). A critical insight of this study is that Bayesian optimization cannot directly deal with categorical parameters, processor-specific parameter constraints, and the slow design evaluation procedure with FPGA synthesis. The key novelty is that we modify the mathematical form of BO to cope with these shortcomings rather than using BO directly. Compared with existing BO approaches for soft processors, ASOP involves the following new components: (i) customized BO covariance kernel to consider categorical parameters, (ii) smooth functions to encode parameter constraints, (iii) speed-aware BO acquisition function to speed up evaluation.

%The platform comprises two primary components. The first component is a parameter optimizer based on BO for processor design, which is highly effective in navigating vast and complex design spaces with limited data samples~\cite{Take_human_out_of_loop, Tutorial_of_Bayesian}. The second component is an automated framework that performs fast evaluations of soft processors across various configurations.

The development of ASPO involves addressing several key challenges. 
\begin{itemize}
    \item \textbf{Challenge C1:} The design space of FPGA-based soft processors includes numerous categorical parameters, including types of branch predictors and division circuits. These categorical parameters are particularly hard to optimize because they lack a natural order or continuity, making them incompatible with standard BO methods~\cite{Dealing_with_categorical} and conventional constraint-handling techniques.

    \item \textbf{Challenge C2:} Various types of constraints need to be imposed among the design parameters~\cite{Boom_Explorer}. Allowing designers to specify these constraints is particularly valuable in FPGA designs, where strict resource limitations and architectural requirements must be adhered to. While these constraints are typically encoded as functions with Boolean outputs, standard BO methods require constraint functions with continuous and smooth ranges, making direct integration of Boolean constraints difficult.

    \item \textbf{Challenge C3:} The optimization process requires evaluating numerous design parameter configurations, which is particularly time-consuming due to the need for accurate evaluation through FPGA synthesis. This exhaustive process significantly increases computational overhead, especially when configurations with minimal variations are repeatedly generated and evaluated.
\end{itemize}

The key contributions of this work are outlined as follows:
\begin{itemize}
    \item A BO workflow to account for the properties of categorical parameters, enhancing the efficiency of the optimization process, addressing Challenge C1. This technique is discussed in Section~\ref{sec:Param_Optimise}.

    \item An innovative method involving the disjunctive form to enable BO to identify all types of constraints during its generation phase, ensuring that the resulting soft processor configuration complies with its specifications, addressing Challenge C2. This method is elaborated in Section~\ref{sec:smooth}.
    
    \item A checkpoint-based method to accelerate the evaluation process by modifying the BO acquisition function to consider previously tested processor configurations. This approach addresses Challenge C3 and is detailed in Section~\ref{sec:framework}, as shown in Figure~\ref{fig:proposed_framework}.
\end{itemize}

The ASPO optimization platform, comprising the Bayesian optimizer, evaluation framework, and scripts for replicating the experiments presented in this paper, is made available as open-source software\footnote{Web address removed for blind review.}.

\section{Background}
\label{sec:background}

Optimizing soft processors has become increasingly reliant on black-box techniques like Bayesian optimization (BO)~\cite{Boom_Explorer, Boomerang}. This section outlines the core principles and methodologies underpinning these optimization strategies.

Soft processor optimization is a black-box optimization problem, characterized by intricate and often opaque relationships between design parameters, processing speed, power consumption, and resource utilization. The absence of explicit mathematical models and gradient information for the objective function further complicates the process, making gradient-based approaches unsuitable. Instead, designers must rely on simulations or direct empirical testing to evaluate configurations, a process that is both resource-intensive and time-consuming.

BO \cite{shahriari2015taking} addresses these challenges by employing a probabilistic surrogate model to approximate the behavior of the objective function. Suppose the design parameter configuration is denoted as $\dpv{x}$, and $f(\dpv{x})$ represents a performance metric such as wall-clock execution time, latency, or clock frequency. A Gaussian process (GP) \cite{williams2006gaussian} is typically used as the surrogate, offering probabilistic predictions of $f(\dpv{x})$. Decisions about which configuration to evaluate next are guided by an acquisition function, balancing the exploration of uncertain regions and the exploitation of promising areas. This iterative process continues until pre-defined criteria, such as convergence or a maximum number of iterations, are met.

\begin{table*}[t]
    \centering
    \caption{Bayesian Optimization Techniques for Tuning Processor Design Parameters}
    \label{tab:bo}
    \scalebox{0.9}{
    \begin{tabular}{llllllll}
    \hline
        Method & FIST & BE & Boomerang & Orbit-ML & VBO & RCBO & ASPO \\ 
        Reference & \cite{xie2020fist} & \cite{Boom_Explorer} & \cite{Boomerang} & \cite{coutinho2023exploring} & \cite{gao2024vanilla} & \cite{my_work} & This paper \\ 
        Year & 2020 & 2021 & 2023 & 2023 & 2024 & 2024 & 2025 \\ 
    \hline
        Target platform & ASIC & ASIC & ASIC & Generic & ASIC & FPGA & FPGA \\ 
        Surrogate model & XGBoost & GP & GP & GP & GP & GP & \textbf{Enhanced GP for categorical parameters} \\
        Acquisition function & Standard & Standard & Standard & Standard & Standard & Standard & \textbf{Penalized with FPGA synthesis speed} \\
        Number of out-of-box processors & 1 & 1 & 1 & 1 & 2 & 1 & \textbf{3} \\ 
        Application-specific optimization & No & No & No & \textbf{Yes} & No & \textbf{Yes} & \textbf{Yes} \\
        \hline 
        Parameter constraint awareness & No & No & No & No & No & No & \textbf{Yes} (Section~\ref{sec:constraints}) \\ 
        Interface for arbitrary FPGA-based processor & \textbf{Yes} & No & No & \textbf{Yes} & No & No & \textbf{Yes} (Section~\ref{sec:framework}, illustrated in Fig.~\ref{fig:proposed_framework}) \\
        Accelerated design evaluation for FPGA & No & No & No & No & No & No & \textbf{Yes} (Section~\ref{sec:framework}, illustrated in Fig.~\ref{fig:Accelerated_Flow}) \\         FPGA resource awareness & No & No & No & No & No & \textbf{Yes} & \textbf{Yes} \\ 
        Performance evaluation with physical layout  & No & No & \textbf{Yes} & No & No & No & \textbf{Yes} (Section~\ref{sec:eval:overall}) \\ 
        \hline
        Open-source software & No & \textbf{Yes} & No & \textbf{Yes} & No & \textbf{Yes} & \textbf{Yes} \\ 
    \hline
    \end{tabular}
    }
\end{table*}

\begin{figure*}[tbp]
  \centering
  \includegraphics[width=\textwidth]{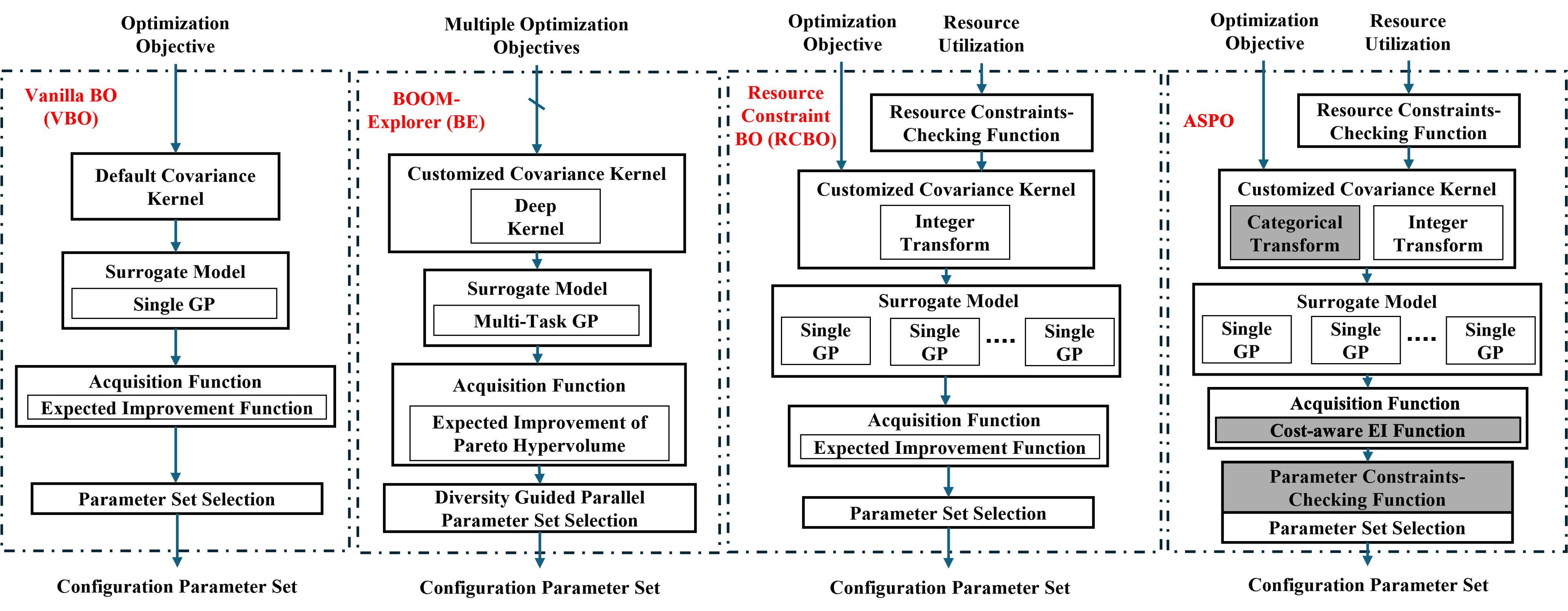}
  \caption{This comparison evaluates the workflows of ASPO, RCBO~\cite{my_work}, BOOM-Explorer~\cite{Boom_Explorer}, and vanilla BO~\cite{gao2024vanilla}. Compared to RCBO, ASPO incorporates an additional categorical transformation within the customized covariance kernel, optimizing its handling of categorical parameters. Additionally, ASPO introduces a parameter constraints-checking function to address inter-parameter constraints. In contrast, BOOM-Explorer is designed for multi-objective scenarios but does not account for resource constraints or parameter dependencies and is not optimized for effectively handling categorical and integer parameters. Lastly, ASPO also introduces a cost-aware Expected Improvement (EI) acquisition function, enhancing the efficiency of configuration selection during the BO process. }
  \label{fig:Proposed_BO}
\end{figure*}

BO has demonstrated significant potential in optimizing processor design parameters. Table~\ref{tab:bo} summarizes and compares the key features of Bayesian optimization methods for tuning processor designs over the past five years, including FIST~\cite{xie2020fist}, BOOM-Explorer (BE)~\cite{Boom_Explorer}, Boomerang~\cite{Boomerang}, Orbit-ML~\cite{coutinho2023exploring}, RCBO~\cite{my_work}, and our proposed method. Furthermore, the workflows of four representative methods are illustrated in Figure~\ref{fig:Proposed_BO}.

Besides processor optimization, BO has been applied beyond architectural parameter tuning to various aspects of hardware and software optimization on FPGA platforms. For instance, BO has been employed to refine computational modules, such as in \cite{li2023amg}, where it is used to generate approximate multipliers that balance accuracy and hardware efficiency. Similarly, \cite{nguyen2020layer} demonstrates the application of BO to determine optimal layer sparsity in FPGA-based object detectors, showcasing its adaptability across domains. Additionally, BO has been leveraged for automation at multiple levels of FPGA design, including routing \cite{zheng2022graebo} and High-Level Synthesis (HLS) code optimization \cite{kuang2023multi, kuang2023hgbo}.
\section{Integrating Categorical Parameters and Parameter Constraints into Optimization}
\label{sec:constraints}
% The parameter constraints are defined by the soft processor’s design specifications and are set within the design space. 

Standard Bayesian Optimization (BO) methods for FPGA design do not support the consideration of parameter constraints defined in the soft processor's design specifications. Consequently, the optimizer may produce invalid designs that fail during configuration, simulation, or FPGA synthesis, or exceed resource limitations. Evaluating these invalid designs reduces productivity and wastes computational resources. Examples of such parameter constraints are presented in Table~\ref{tab:boom_constraints}.

To address this, we propose three types of parameter constraint-checking functions, which are used to examine the validity of configuration parameter sets against three classes of constraints. These functions are integrated into the parameter selection process of the BO method. This integration allows the BO method to handle constrained optimization problems by maximizing the acquisition function while satisfying feasibility requirements~\cite{Tutorial_of_Bayesian}.

Under this framework, gradient-based constrained optimization solvers, such as Sequential Least Squares Programming (SLSQP)~\cite{SLSQP} and trust-region algorithms~\cite{trust_region}, can be employed alongside the optimizer. However, the application of these methods has two key requirements. First, the domain of the constraint functions must be numeric. Second, the range of the functions must be smooth to effectively guide the search for feasible and optimal solutions.

In typical FPGA-based soft processor designs, parameter sets often include categorical design parameters, which make the domain non-numeric. Moreover, traditional constraint functions output Boolean values (true or false), indicating whether a constraint is met, but these outputs lack the smoothness required for gradient-based optimization. The following two subsections address these challenges by adapting the BO covariance kernel to better handle categorical parameters and formulating constraint functions with smooth numerical ranges, respectively.

\subsection{Adaptation for Categorical Parameters}
\label{sec:Param_Optimise}

In ASPO, a one-hot encoding method is introduced to encode all categorical related design parameters into a vector, where each element corresponds to a specific category. During the BO process, every element in this vector is treated as an independent real-valued variable bounded within \([0, 1]\). This transformation enables categorical parameters to be processed within a structured, high-dimensional continuous space that supports gradient-based optimization. Additionally, it explicitly captures the dependencies between categorical and real-valued parameters, allowing models to better learn and predict performance based on these interactions.

Additionally, the covariance kernel \(K(\mathbf{x}, \mathbf{x}')\) in BO is customized to accommodate these categorical parameters. This customized kernel is essential for calculating the covariance function \(\sigma^2(\mathbf{x}, \mathbf{x}')\), which measures the similarity or correlation between different sample points in the input space. The covariance function guides the acquisition function in selecting the next point to sample~\cite{Take_human_out_of_loop}. 

\begin{table*}[htbp]
    \centering
    \caption{Predefined Design Space for RocketChip and BOOM Processors in the Experiment}
    \label{Tab:rocketchip_design_space}
    \footnotesize
    \scalebox{0.9}{%
    \begin{minipage}{\textwidth}
    \centering
        \begin{tabular}{p{1.4cm}p{1.7cm}p{8cm}p{4cm}p{1.7cm}}
            \hline
            Processor & Parameter & Description & Values & Default Config\\
            \hline
            \multirow{5}{*}{EL2 VeeR}
            & icache size
            & Instruction Cache Size in KB
            & \{16, 32, 64, 128, 256\} & 16\\
            & lsu\_stbuf\_depth
            & Number of store operations the LSU buffer can hold simultaneously
            & \{2, 4, 8\} & 4\\
            & btb\_enable
            & Boolean parameter deciding whether to include BTB.
            & \{True, False\} & True\\
            & iccm\_size 
            & Instruction closely coupled memory size
            & \{4, 8, 16, \ldots, 512\} & 64 \\
            & dccm\_size 
            & Data closely coupled memory size
            & \{4, 8, 16, \ldots, 512\} & 64\\
            \hline
            \multirow{5}{*}{\shortstack{RocketChip \\ \& BOOM}}
            & core\_num 
            & The number of cores. 
            & \{1, 2, 3, 4\} & 1\\
            & icache\_nSets 
            & Number of sets in the set-associative icache.
            & \{2, 4, 8, 16, 32, 64\} & 64 \\
            & icache\_nWays 
            & Number of ways in each set of the icache. 
            & \{2, 4, 8, 16, 32, 64\} & 4 \\
            & dcache\_nSets 
            & Number of sets in the set-associative dcache. 
            & \{2, 4, 8, 16, 32, 64\} & 64\\
            & dcache\_nWays 
            & Number of ways in each set of the dcache. 
            & \{2, 4, 8, 16, 32, 64\} & 4 \\
            \hline
            \multirow{2}{*}{RocketChip}
            & mul\_div\_config 
            & Configuration of the multiplication and division unit. 
            & \{\texttt{Fast}, \texttt{Default}, \texttt{Simple}\} & \texttt{Default}\\
            & btb\_config 
            & Configuration of the branch target buffer. 
            & \{\texttt{Default}, \texttt{WithoutBTB}\} & \texttt{Default} \\
            \hline
            \multirow{5}{*}{BOOM}
            & bpd\_config
            & Configuration of the branch predictor.
            & \{\texttt{TAGEL}, \texttt{Boom2}, \texttt{Alpha21264}\} & \texttt{TAGEL}\\
            & FetchWidth 
            & The number of instructions the fetch unit can retrieve per cycle. 
            & \{1, 4, 8\} & 4\\
            & DecodeWidth 
            & The number of instructions the decode unit can process simultaneously. 
            & \{1, 2, 3, 4, 5, 6\} & 1\\
            & RobEntry 
            & The number of reorder buffer entries. 
            & \{32, 64, 96, 128, 120\} & 32 \\
            & FetchBufferEntry 
            & The number of entries in the instruction fetch buffer. 
            & \{8, 16, 24, 32, 35, 40\} & 16 \\
            \hline
        \end{tabular}
    \end{minipage}
    }
\end{table*}
The kernel customization is to introduce a categorical transformation for the one-hot encoded vectors derived from categorical parameters before the covariance kernel computation. This transformation sets the maximum value within each one-hot encoded vector to one and all remaining elements to zero. As a result, samples that map to the same one-hot vectors share the same customized kernel function value \(K'(\mathbf{x}, \mathbf{x}')\). Consequently, when a sample is evaluated, the covariance function \(\sigma^2(\mathbf{x}, \mathbf{x}')\) for any sample that can be transformed to the same one-hot vector becomes zero~\cite{my_work}. This causes the acquisition function to exclude these regions from further sampling, thereby effectively preventing the repetitive selection of identical soft processor designs and enhancing the efficiency of the BO process.

\subsection{Constraint Functions with Smooth Numeric Ranges}
\label{sec:smooth}
There are three main types of parameter constraint, either from the specification~\cite{Boom_Explorer} or from the empirical experiment. Some example constraints for the BOOM processor are listed in Table~\ref{tab:boom_constraints}.

\begin{table}[h]
    \centering
    \caption{Example Constraints of BOOM design specification}
    \begin{tabular}{c l}
       \hline
        Classification & Descriptions \\ 
        \hline
        \multirow{2}{*}{Inequalities}
         & FetchWidth $\geq$ DecodeWidth \\ 
         & FetchBufferEntry $>$ FetchWidth \\ 
         \hline
         \multirow{3}{*}{Conditional}
         & if \(\text{icache\_nWays} \in [64, 128]\), then \(\text{nSets} \in [2, 4]\) \\
         & \{ if \(\text{dcache\_nWays} \in [16, 32]\), then \(\text{nSets} \in [2, 4]\); \\
         & or if \(\text{dcache\_nWays} \in [128, 256]\), then \(\text{nSets} \in [4, 8]\) \} \\
         \hline
         \multirow{2}{*}{Divisibility}
         & RobEntry $\mid$ DecodeWidth \\
         & FetchBufferEntry $\mid$ DecodeWidth \\
        \hline
    \end{tabular}
    \label{tab:boom_constraints}
\end{table}

To formulate these three constraint-checking functions, we define that a negative output from either function signifies unmet constraints, while a non-negative output indicates that the constraints are satisfied. For the inequality constraint-checking function, denoted as \(PC_{inequality}(x)\), all inequality constraints are generalized into the following format:
\begin{equation}
    PC_{inequality}(x) = k_a x_a - k_b x_b + t
    \label{Eq:Parameter_inequality_Constraints}
\end{equation}
where \(k_a\), \(k_b\), and \(t\) are real-valued parameters that are automatically assigned based on inequality constraints derived from the processor's specifications.

In addition to inequality constraints, ASPO also supports non-linear conditional constraints. Their relationships can be categorized as either conjunctive or disjunctive. A conjunctive structure consists of multiple conditional constraints and is satisfied only when all constraints are met. In contrast, a disjunctive structure is satisfied if at least one of its conditional constraints is true.

With minimal human effort, all the conditional constraints in the processor's specification can be systematically organized into a hierarchical logical structure comprising conjunctive and disjunctive components. At the top level, a conjunctive constraint combines all the underlying conditional constraint structures. This hierarchical organization serves as the target for our proposed constraint-checking function, \(PC_{cond}(x)\), which evaluates the constraints as follows:

% At the lowest level, each building block of this structure is also a conjunctive constraint. 
\begin{center}
\begin{tikzpicture}[
    every node/.style={font=\normalsize},
    constraint/.style={draw=red, thick, ellipse, inner sep=1pt},
    label/.style={red, font=\scriptsize, anchor=north},
    arrow/.style={->, thick}
]

% Main structure
\node (PCcond) at (0, 0) {$PC_{cond}$};
\node[constraint] (Cconj1) at (2, 0) {$C_{conj}$};
\draw[arrow] (PCcond) -- (Cconj1);

% Top-level label
\node[label] at (2, -0.5) {Top Level};

% Disjunctive branch
\node at (3, 0) {$\Big\{$};
\node (Cdisj) at (3.7, 0.3) {$C_{disj}$};
\node at (3.7, 0) {$\vdots$};
\node (cconj1) at (3.7, -0.45) {$C_{conj}$};
\node (cbb) at (4.9, -0.45) {$C_b$};
\draw[arrow] (cconj1) -- (cbb);

% Building blocks
\node (Cbb1) at (5.45, 0.52) {$C_b$};
\draw[arrow] (Cdisj) -- (4.5, 0.3);
\node (Cbb2) at (5.45, 0) {$C_b$};

\node at (4.8, 0.3) {$\Big\{$};

\end{tikzpicture}
\end{center}

The single basic conditional constraint example format and the corresponding constraint-checking function, \(C_{b}\) are illustrated below:

\begin{itemize}
    \item If \(x_1\) lies within the interval \([a_{1}, b_{1}]\), then \(x_2\) must lie within the interval \([a_{2}, b_{2}]\).
\end{itemize}

\begin{align}
    C_{b}(x) &= \min_{m} c_{m}(x_m), \quad \forall m \in \{1, \ldots, d\}
    \label{Eq:Disjunctive_Conditional_Constraints}
    \\
    c_{m}(x_m) &= - (x_m - a_{m})(x_m - b_{m})
\end{align}

Here, each \(c_{m}(x_m)\) is differentiable, ensuring sufficient smoothness to support the Bayesian optimizer.

For a disjunctive structure of constraints, the constraint-checking function, \(C_{disj}\), is defined as:

\begin{equation}
    C_{disj} = \max_i C_i(x), \quad \forall i \in \{1, \ldots, k\},\
\end{equation}
where each \(C_i(x)\) represents the \( i \)-th conditional constraint of the \(k\) conditional constraints of the disjunctive structure. It can be \(C_{b}\), \(C_{conj}\), or \(C_{disj}\). Given that these functions are defined such that a nonnegative output indicates that the input \(x\) satisfies the evaluated constraint, taking the maximum value in \(C_{disj}\) ensures that if at least one \(C_i(x)\) returns a nonnegative value, the disjunctive constraint structure is satisfied.

Conversely, for a conjunctive structure of constraints, the constraint-checking function, \(C_{conj}\), is determined by taking the minimum value among all conditional constraints within the structure:

\begin{equation}
    C_{conj} = \min_i C_i(x), \quad \forall i \in \{1, \ldots, k\}
\end{equation}

This approach ensures that if the smallest output is negative, at least one of the conditional constraints is violated, indicating that the conjunctive constraint is not satisfied.

For the third type of constraint, which involves checking the divisibility between parameters, we propose the following checking function:

\begin{equation}
    PC_{\text{divisibility}}(x) = -\sin^2\left(\frac{\pi x_a}{x_b}\right)
\end{equation}

The function \( PC_{\text{divisibility}}(x) \) returns zero only when \( x_b \) is divisible by \( x_a \); otherwise, it returns a negative value. The use of a trigonometric function ensures differentiability for the BO optimizer, while its periodic nature guarantees that the function's value depends solely on whether \( \frac{x_a}{x_b} \) is an integer and not on its magnitude.

\begin{figure}[tbp]  % The placement specifier can be [htbp]
  \centering
  \includegraphics[width=0.48\textwidth, trim=0 1cm 0 0]{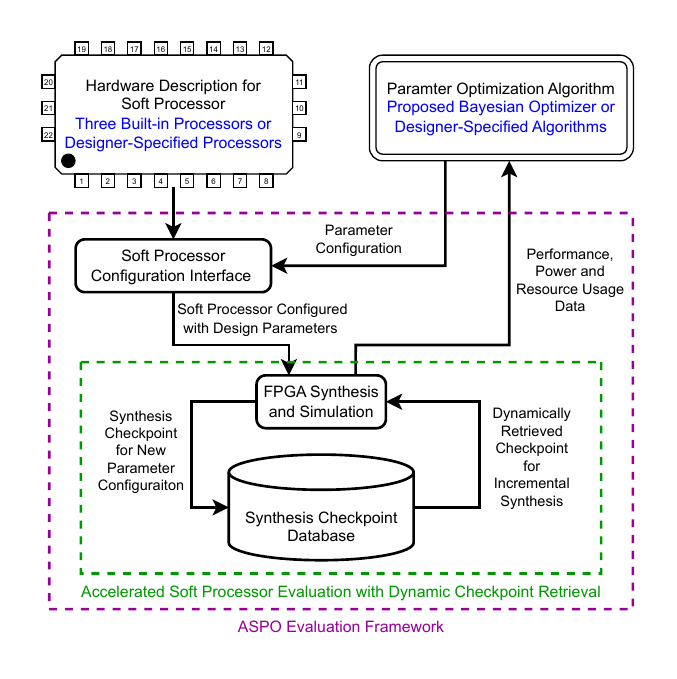}
  \caption{Overview of the proposed framework for accelerated soft processor design and evaluation. The framework supports three built-in processors or designer-specified processors, which can be configured using a parameter optimization algorithm, such as the proposed Bayesian optimizer or alternative designer-specified methods. Design parameters are iteratively optimized to achieve desired performance objectives. The evaluation framework integrates FPGA synthesis and simulation tools, using a dynamic checkpoint retrieval system to reuse synthesis data from previously evaluated configurations, significantly accelerating the evaluation process.}
  \label{fig:proposed_framework}
\end{figure}

\section{Accelerated Soft Processor Evaluation}

\label{sec:framework}
The primary goal of our soft processor evaluation framework is to provide a rapid and automated assessment of the performance of specified soft processor configurations. Since the FPGA synthesis process is often a significant bottleneck in the optimization workflow, this framework adopts a data-driven approach to accelerate the evaluation process. The detailed workflow is illustrated in Figure~\ref{fig:proposed_framework}.

\subsection{Checkpoint Retrieval for Fast Design Evaluation}
\label{subsec: Evaluation Acceleration}

During the processor design process, design methods often produce configurations that share several components with previously evaluated processors. This overlap allows us to store key information from the evaluation of earlier configurations and reuse it to expedite the evaluation of new designs. Modern synthesis tools, such as Vivado and Quartus Prime, support incremental synthesis, which leverages stored synthesis data in checkpoint files to streamline subsequent runs~\cite{Incremental_Synthesis}. When synthesizing a new configuration that differs from a previously synthesized version, these tools detect changes and re-synthesize only the modified components, significantly reducing synthesis time.

\begin{figure}[htbp]  % The placement specifier can be [htbp]
  \centering
  \includegraphics[width=0.5\textwidth]{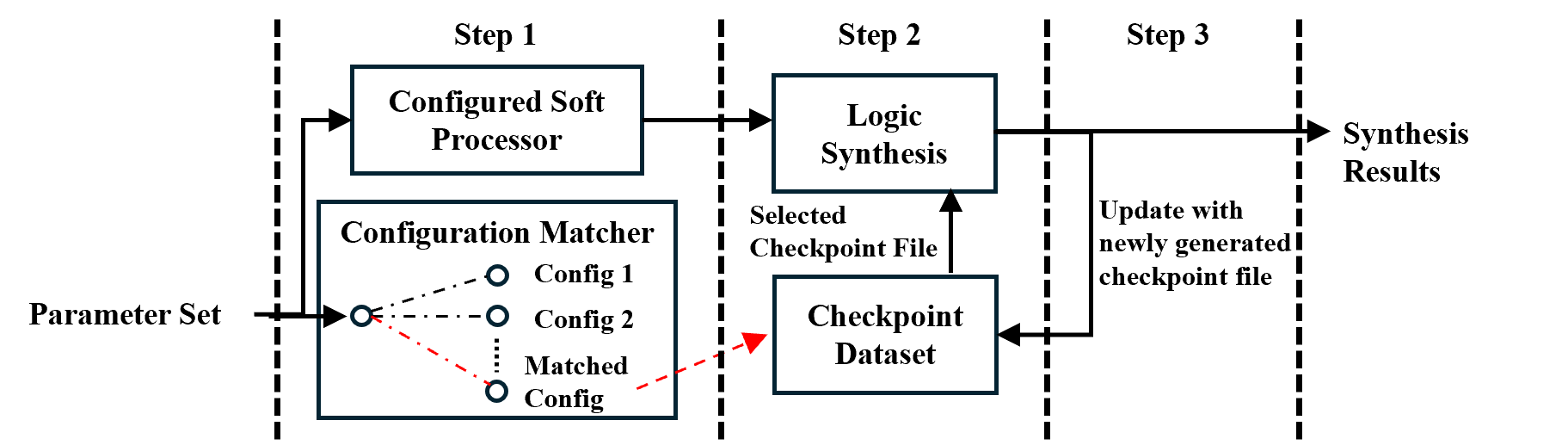}
  \caption{Accelerated evaluation flow.}
  \label{fig:Accelerated_Flow}  % Label for referencing the figure in the text
\end{figure}

To take advantage of the incremental synthesis feature for faster evaluation, two key components are required. The first is a database of checkpoints to store designs that can serve as starting points for incrementally synthesizing new configurations. The second is a matching function to identify an appropriate starting point from the database, minimizing potential synthesis time by reusing previously synthesized configurations.

We propose dynamically building the checkpoint database during the optimization process. This database stores checkpoint files generated from previously synthesized processor configurations, allowing efficient reuse in subsequent synthesis tasks. For reference checkpoint selection, we introduce a configuration matcher that identifies the checkpoint file with the highest overlap with the processor currently under evaluation. By minimizing the number of modified components requiring re-synthesis, this approach significantly reduces synthesis time. The flow of the accelerated evaluation process is illustrated in Figure~\ref{fig:Accelerated_Flow}.

To implement the reference checkpoint selection, we develop a configuration matching function that identifies the most similar configuration in the dataset based on a weighted Euclidean distance metric. The mathematical formulation of the matching function is as follows:

\begin{equation}
     \text{MatchConfig}(\dpv{x},\dpv{w},Q) = \operatorname*{argmin}_{\dpv{q} \in Q} \sum_{i=0}^{d-1} w_{i} \cdot (x_{i} - q_{ni})^2
    \label{eq:config_matcher}
\end{equation}

In this equation, \( x_i \) represents the \( i \)-th parameter of the configuration \(\dpv{x}\) being evaluated, and \( q_{ni} \) denotes the \( i \)-th parameter of the \( n \)-th stored configuration \(\dpv{q}\) within the checkpoint dataset \( Q \). The weight vector \(\dpv{w} \in \mathbb{R}^d\) quantifies the relative importance of changes in each parameter for configuration matching. We aim to compute the optimal weights \( \mathbf{w}^* \) by solving the following optimization problem:
\begin{equation}
    \mathbf{w^*} = \operatorname*{argmin}_{ \dpv{w} \in \mathbb{R}^{d}} \sum_{\dpv{q} \in Q} \mathcal{T}_\mathrm{syn}(\dpv{q}, \text{MatchConfig}(\dpv{q}, \dpv{w}, Q\setminus \dpv{q}))
    \label{eq:weight_optimization}
\end{equation}

Here, \(\mathcal{T}_\mathrm{syn}(\dpv{x}, \dpv{y})\) represents the synthesis time required to process configuration \(\dpv{x}\) using the checkpoint from configuration \(\dpv{y}\). This optimization determines the optimal weight vector \(\mathbf{w^*}\) by minimizing the total synthesis time across all configurations in the dataset \(Q\). Given the substantial size of the dataset and the lengthy synthesis time for each configuration, an approximate solution is obtained by randomly sampling a small subset of design configurations and applying heuristic search techniques.

\subsection{Warm Start}

The evaluation process in our proposed framework becomes increasingly efficient as more soft processor configurations are evaluated. With each evaluated configuration, the checkpoint database grows with an additional checkpoint file. This expansion enables the configuration matcher to identify checkpoint files from previously evaluated configurations with greater overlap to the current configuration being tested, thereby accelerating the evaluation process. Additionally, a larger database increases the likelihood that the performance of a new configuration has already been recorded, eliminating the need for re-evaluation.

To achieve high evaluation speeds, implementing a warm-start feature that strategically guides the sampling of processor configurations is essential. This feature ensures that sampled configurations are uniformly distributed across the entire design space, maximizing the chances that any new configuration to be evaluated will have a closely matched configuration in the database, thus improving evaluation efficiency.

Traditional sampling strategies, such as random sampling, are avoided due to their inefficiency in achieving uniform distribution across the design space. This inefficiency arises because the design space is vast and consists of multiple parameters, each with its own distinct range of possible values. To address this challenge, the orthogonal array (OA) strategy is adopted. The orthogonality of OA ensures that sampled configurations are evenly distributed throughout the design space~\cite{OA}. Since the design space for the soft processor is fixed within our framework, the OA can be pre-generated using online methods, further simplifying implementation.

\subsection{Incorporating Evaluation Time Cost Into Optimization}
Due to the proposed accelerated soft processor evaluation method, the evaluation time of a configuration depends on how similar it is to previously evaluated configurations stored in the checkpoint database. Therefore, we incorporate the consideration of evaluation time in the BO to enable more efficient configuration selection and reduce the total evaluation time.

To achieve this, we incorporate a cost-aware cooling mechanism into the BO acquisition function, steering configuration selection in the BO process for optimized overall evaluation time. The cost function, $\hat{c}(x)$, is defined as the minimum weighted Euclidean distance between the candidate configuration and all previously evaluated configurations stored in the checkpoint database, same as Equation~\ref{eq:config_matcher}. Inspired by~\cite{costaware_bo}, the modified cost-aware acquisition function is defined as:

\begin{equation}
    \alpha_{\text{cool}}(x, t) = \frac{\alpha(x)}{\lambda(t) \cdot \hat{c}(x)}
\end{equation}

where $\alpha(x)$ is the original acquisition function (Expected Improvement in this project), and $\lambda(t)$ denotes a cooling schedule that evolves over the current BO iteration $t$. In this work, we adopt a simple decaying schedule designed to impose a high penalty on distant configurations in the early optimization stages. This encourages the algorithm to remain in regions that are quick to evaluate. As the optimization progresses, the penalty is gradually reduced, allowing the algorithm to explore more distant, potentially expensive but promising configurations. The cooling schedule is defined as:

\begin{equation}
    \lambda(t) = \lambda_0 \cdot \exp(-k t)
\end{equation}

where $\lambda_0$ is the initial cost sensitivity, and $k$ is the decay rate; both are determined empirically through experiments.

\begin{table}[htbp]
    \centering
    \caption{RISC-V Benchmarks used in this project.}
    \label{Tab:Benchmarks}
    \begin{tabular}{c l c}
        \hline
        Benchmark & Focus & minstret \\
        \hline
        coremark~\cite{coremark} & Basic operations & 282995 \\
        dhrystone~\cite{dhrystone} & Integer arithmetic and string handling & 186031 \\ 
        rsort~\cite{newsome2019riscvtests} & Memory access & 171154 \\
        qsort~\cite{newsome2019riscvtests} & Branching, recursion and memory access & 123506 \\
        multiply~\cite{newsome2019riscvtests} & Multiplication & 42503 \\
        spmv~\cite{newsome2019riscvtests} & Floating point and memory access & 34466 \\
        mm~\cite{newsome2019riscvtests} & Cache and memory access & 24744 \\
        % mt-memcpy~\cite{newsome2019riscvtests} & Cache and memory access & 14674 \\
        % median & Applies a three-element median filter to data, evaluating the processor's data-cache performance. & 4659 \\
        % vadd & Assesses the processor's performance in vector addition. & 2416 \\
        \hline
    \end{tabular}
\end{table}

% \begin{table*}[htbp]
% \centering
% \caption{Performance Metrics Recorded in the Dataset}
% \label{tab:soft_processors}
% \begin{tabular}{c c c }
% \hline
% \textbf{Type} & \textbf{Metric} & \textbf{Description} \\ 
% \hline
% \multirow{2}{*}{Benchmark Performance} & mcycles & Cycles required by the processor to execute compiled tasks. \\ 
% % \cline{2-3}
%                                        & minstret & Instructions retired or completed by the processor. \\ 
% \hline
% \multirow{3}{*}{Resource Utilization} & LUT Elements & Logic building blocks for RTL designs on FPGA. \\ 
% % \cline{2-3}
%                                        & FFs & Storage elements in sequential logic circuits on FPGA.\\ 
% % \cline{2-3}
%                                        & BRAM & On-chip memory for storing large data amounts on FPGA. \\ 
% \hline
% Power Performance & On-Chip Power & Combined static and dynamic power consumed by the FPGA.\\ 
% \hline
% \multirow{2}{*}{Timing Performance} & Worst Setup Slack & Largest negative deviation from required arrival time of signals. \\ 
% % \cline{2-3}
%                                    & Worst Hold Slack & Largest early arrival time of signals beyond minimum hold time. \\ 
% \hline
% \end{tabular}
% \end{table*}

\section{Evaluation}

We conducted two experiments to assess the performance of ASPO. The first experiment aims to evaluate the acceleration improvements offered by the automated soft processor evaluation framework in Section~\ref{sec:framework}. The second experiment focused on evaluating the overall performance of our proposed soft processor design platform that combines all features in Sections~\ref{sec:constraints} and \ref{sec:framework}.

\subsection{Experiment Setup}

Our experimental implementation of the proposed framework supports three soft processors out-of-box: RocketChip~\cite{RocketChip}, BOOM~\cite{BOOM}, and EL2 VeeR~\cite{CoresVeeREL2}. These processors were selected for their high performance, extensive design flexibility, and support for FPGA deployment. With minor modifications, the implementation can be extended to support other configurable RISC-V soft processors, such as Ibex~\cite{ibex_core} and CVA6~\cite{CVA6}.

All experiments were conducted on a desktop workstation equipped with an Intel Xeon E5-1650 v3 CPU and 32GB of DDR4 RAM, without utilizing a GPU. GPU acceleration was deemed unnecessary, as the Bayesian optimizer contributes only a small portion of the total runtime and would not benefit significantly from GPU usage. The evaluation phase, primarily involving processor simulation using Verilator and FPGA hardware synthesis using Vivado, is the most time-consuming step and cannot be accelerated by a GPU.

The evaluation includes simulation results for multiple RISC-V benchmarks supported by soft processors, such as dhrystone and coremark, as listed in Table~\ref{Tab:Benchmarks}. These benchmarks are designed to evaluate the processor's performance across various processing aspects. The complexity of each benchmark is measured by the machine instructions retired (minstret)~\cite{minstret_ref}, representing the total number of instructions executed by the processor. 

The results for resource utilization, power, and timing metrics are derived from the report generated from the synthesis process. The timing metric is assessed by manually configuring the processor's clock input to a default reference frequency of \( 50 \, \text{MHz} \) and extracting the worst setup slack and worst hold slack values from the Vivado timing report. These two values offer insights into the timing performance of the soft processor at this default frequency and can be used to estimate its maximum operating frequency.

\subsection{Experiment for Soft Processor's Evaluation Framework}

This experiment primarily focuses on determining the time required to evaluate a set of soft processor configurations and comparing it with other reference evaluation frameworks. The evaluation time is defined as the duration from receiving the soft processor configuration parameters to the completion of recording all performance metrics, including execution time, resource utilization, power consumption and timing performance. Three design evaluation frameworks are considered:

\textbf{Direct evaluation without Acceleration:} This framework represents the typical evaluation flow for design methods such as RCBO~\cite{my_work}, which assesses each parameter configuration as a separate task and excludes design checkpoints to accelerate the synthesis process.

\textbf{Evaluation with a fixed checkpoint:} This framework sets the checkpoint file used in logic synthesis as the one obtained from the evaluation of the default soft processor configuration and disables the configuration matcher and checkpoint dataset. The default configuration for each soft processor is provided in Table~\ref{Tab:rocketchip_design_space}.

\textbf{Evaluation with our retrieval strategy proposed in Section~\ref{sec:framework}:} For the proposed evaluation framework, we instruct it to complete the automated design space exploration process before the experiment. This ensures that the configuration matcher has a sufficient pool of configuration candidates to select from, thereby enhancing evaluation acceleration.

\begin{figure}
    \centering
    \includegraphics[width=1\linewidth]{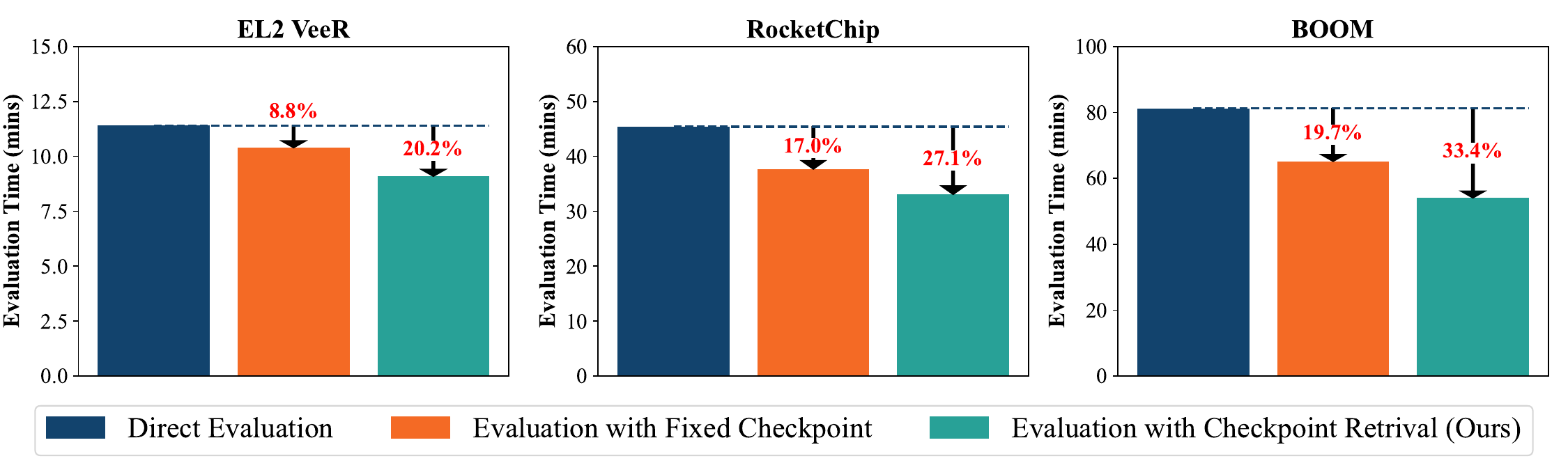}
    \caption{The two frameworks employing acceleration during the logic synthesis stage can significantly enhance evaluation speed by reusing existing evaluation data. Additionally, our proposed framework achieves faster evaluations through a strategic selection of checkpoint files, in contrast to frameworks that consistently utilize the same checkpoint file.}
    \label{fig:framework_evaluation_experiment}
\end{figure}

To ensure a fair comparison, all three evaluation frameworks are provided with an identical set of soft processor configurations. In this experiment, ten randomly generated configurations were created for each supported soft processor, serving as the candidate experiment set. Experimental results are presented in Figure~\ref{fig:framework_evaluation_experiment}, and the quantitative comparison among the three frameworks is shown in Table~\ref{tab:numerical_eval_time}.

\begin{table}[htbp]
    \centering
    \footnotesize
    \renewcommand{\arraystretch}{0.9}
    \caption{Evaluation Time for Design Evaluation (in minutes)}
    \label{tab:numerical_eval_time}
    \scalebox{0.85}{
    \begin{tabular}{l l ccc}
        \toprule
        \textbf{Processor} & \textbf{Evaluation Method} & \textbf{Avg} & \textbf{Min} & \textbf{Max} \\
        \midrule
        \multirow{3}{*}{EL2 VeeR} 
        & Direct Evaluation              & 11.4 & 11.2 & 11.5 \\
        & Evaluation with fixed checkpoint & 10.4 & 8.5  & 11.1 \\
        & Our framework                           & 9.1  & 8.4  & 10.1 \\
        \midrule
        \multirow{3}{*}{RocketChip} 
        & Direct Evaluation              & 45.4 & 44.7 & 46.2 \\
        & Evaluation with fixed checkpoint & 37.7 & 24.0 & 43.9 \\
        & Our framework                           & 33.1 & 18.0 & 37.5 \\
        \midrule
        \multirow{3}{*}{BOOM} 
        & Direct Evaluation              & 81.2 & 80.1 & 81.9 \\
        & Evaluation with fixed checkpoint & 65.2 & 57.5 & 71.2 \\
        & Our framework                     & 54.1 & 42.5 & 63.4 \\
        \bottomrule
    \end{tabular}
    }
\end{table}

\subsection{Experiment for Overall Design Platform}
\label{sec:eval:overall}
To rigorously evaluate the design platform's performance, this experiment outlines a series of application benchmarks. These tasks involve configuring the three supported soft processors within the predefined design space outlined in Table~\ref{Tab:rocketchip_design_space}, optimizing their performance using selected RISC-V benchmarks listed in Table~\ref{Tab:Benchmarks}, and ensuring deployability on the target FPGA boards. The \textit{Zynq UltraScale+ XCZU3CG} FPGA, offering 70{,}560 available LUTs, is selected as the target platform for EL2 VeeR, while the \textit{Zynq UltraScale+ XCZU6CG}, with 214{,}604 available LUTs, is chosen for BOOM.

To compare the performance of our design platform with other design methods, three metrics are proposed, focusing on both design productivity and design quality.

Design productivity is evaluated using two metrics: \textbf{Invalid Design Rate (IDR)} and \textbf{Total Design Time (TDT)}. IDR quantifies the proportion of designs that fail to meet either the processor's parameter constraints or the FPGA's resource constraints during the design process. TDT measures the total time, in hours, required for the design method to converge to the final processor configuration. While valid processors typically undergo longer evaluations, invalid processors may trigger errors at various stages, leading to variability in evaluation times. To ensure practical feasibility, an upper limit of \textbf{35 hours} is set for TDT, and the experiment is terminated upon reaching this time bound.

Design quality is assessed using \textbf{Estimated Execution Time (EET)} \cite{Boomerang},  which serves as the optimization objective in this experiment. EET estimates the processor’s execution time for the designated tasks on FPGA fabric, providing an accurate reflection of the processor's physical layout performance~\cite{Boomerang}. It is computed by dividing the simulated number of execution cycles by the maximum operating frequency, as reported in the timing performance section of the synthesis report.

Most state-of-the-art processor design methods do not account for parameter constraints or resource constraints. Consequently, modifications were necessary to adapt these methods for our experiment. Specifically, if a method does not support parameter constraint awareness or FPGA resource limitations and generates invalid designs, such designs are discarded without being used to update the surrogate model, preventing the optimizer from being misled. Furthermore, due to the significant evaluation time and the large design space, methods requiring extensive evaluations—such as brute-force exhaustive search algorithms, acquisition-free surrogate optimization~\cite{kurek2014automating}, genetic algorithms~\cite{kao2020gamma}, ant colony optimization~\cite{zhang2008extended}, and reinforcement learning~\cite{bai2024towards}—are excluded from consideration. The design methods selected for comparison include \textbf{Hill Climbing (HC)~\cite{Hill_Climbing}}, \textbf{Vanilla Bayesian Optimization (VBO)~\cite{gao2024vanilla}}, \textbf{BOOM-Explorer (BE)~\cite{Boom_Explorer}}, and \textbf{Resource-Constraint Bayesian Optimization (RCBO)~\cite{my_work}}, as discussed in Section~\ref{sec:background}.

% \begin{table}[htbp]
%     \centering
%     \caption{FPGA Resource Specifications}
%     \label{Tab:FPGA_for_Rocket}
%     \begin{tabular}{cccc}
%     \hline
%         Device Name & Part Number &  LUT Elements & Flip-Flops\\ \hline
%         Zynq 7000 & XC7Z012S & 32600  &  65200 \\
%         % Zynq 7000 & XC7Z014S & 40600  &  81200 \\ 
%         Zynq Ultrascale & XCZU3CG & 70560 & 141120 \\
%         % Zynq Ultrascale & XCZU5EV & 117120 & 234240 \\
%         Zynq Ultrascale & XCZU6CG & 214604 & 429208 \\
%         \hline
        
%     \end{tabular}
% \end{table}

\begin{table}[htbp]
\centering
\footnotesize
\renewcommand{\arraystretch}{0.8}
\caption{Design Productivity}
\label{Tab:Experiment_Results_in_Design_Efficiency}
TDT: Total Design Time in Hours ~~ IDR: Invalid Design Rate
\setlength{\tabcolsep}{4.5pt}
\scalebox{0.8}{
\begin{tabular}{@{}cc c ccccc@{}}
\toprule
\textbf{Processor} & 
\textbf{Task} & 
\textbf{Metric} & 
\textbf{HC} & 
\textbf{VBO} & 
\textbf{BE} & 
\textbf{RCBO} & 
\textbf{ASPO} \\
\midrule

%=========================================================
% EL2 VeeR -- coremark
\multirow{18}{*}{\textbf{EL2 VeeR}} 
 & \multirow{2}{*}{coremark} 
 & TDT & 
 0.56$^\dagger$ & 2.61 & 2.18 & 1.93 & \textbf{1.04} \\
 &  
 & IDR & 
\textbf{0\%} & 14\% & \textbf{11\%} & \textbf{0\%} & \textbf{0\%} \\

\cmidrule(lr){2-8}
% EL2 VeeR -- dhrystone
 & \multirow{2}{*}{dhrystone}
 & TDT & 
 0.61$^\dagger$ & 3.98 & 2.12 & 1.88 & \textbf{1.63} \\
 &  
 & IDR & 
 17\% & \textbf{0\%} & 14\% & \textbf{0\%} & 7\% \\

\cmidrule(lr){2-8}
% EL2 VeeR -- rsort
 & \multirow{2}{*}{rsort}
 & TDT & 
 0.81$^\dagger$ & 1.14 & 3.46 & \textbf{0.98} & 1.01 \\
 &
 & IDR & 
 13\% & \textbf{0\%} & 22\% & \textbf{0\%} & \textbf{0\%} \\

\cmidrule(lr){2-8}
% EL2 VeeR -- qsort
 & \multirow{2}{*}{qsort}
 & TDT & 
 0.53$^\dagger$ & 3.43 & 1.33 & 1.02 & \textbf{0.92} \\
 &
 & IDR & 
 \textbf{0\%} & \textbf{0\%} & 12\% & \textbf{0\%} & \textbf{0\%} \\

\cmidrule(lr){2-8}
% EL2 VeeR -- multiply
 & \multirow{2}{*}{multiply}
 & TDT & 
 1.01 & 1.41 & 2.13 & 1.15 & \textbf{0.72} \\
 & 
 & IDR & 
 20\% & \textbf{0\%} & 13\% & 11\% & 14\% \\

\cmidrule(lr){2-8}
% EL2 VeeR -- spmv
 & \multirow{2}{*}{spmv}
 & TDT & 
 0.44$^\dagger$ & 2.65 & 1.50 & \textbf{1.11} & 1.13 \\
 &  
 & IDR & 
 \textbf{0\%} & 18\% & 11\% & \textbf{0\%} & \textbf{0\%} \\

\cmidrule(lr){2-8}
% EL2 VeeR -- mm
 & \multirow{2}{*}{mm}
 & TDT & 
 0.64$^\dagger$ & 1.43 & 1.98 & 1.15 & \textbf{1.08} \\
 & 
 & IDR & 
 40\% & \textbf{0\%} & 18\% & \textbf{0\%} & \textbf{0\%} \\

\cmidrule(lr){1-8}

%=========================================================
% RocketChip -- coremark
\multirow{18}{*}{\textbf{RocketChip}} 
 & \multirow{2}{*}{coremark} 
 & TDT & 
 3.56$^\dagger$ & 27.11 & 22.08 & 18.13 & \textbf{11.24} \\
 &  
 & IDR & 
75\% & 52\% & 38\% & 33\% & \textbf{21\%} \\

\cmidrule(lr){2-8}

% RocketChip -- dhrystone
 & \multirow{2}{*}{dhrystone}
 & TDT & 
 3.00$^\dagger$ & 20.12 & 23.41 & 14.12 & \textbf{11.25}  \\
 & 
 & IDR & 
 88\% & 42\% & 48\% & 37\% & \textbf{31\%} \\

\cmidrule(lr){2-8}

% RocketChip -- rsort
 & \multirow{2}{*}{rsort}
 & TDT & 
 4.00$^\dagger$ & \textbf{8.43} & 16.33 & 15.12 & 10.12 \\
 &
 & IDR & 
 71\% & 40\% & 52\% & 36\% & \textbf{21\%} \\

\cmidrule(lr){2-8}
% RocketChip -- qsort
 & \multirow{2}{*}{qsort}
 & TDT & 
 5.30$^\dagger$ & 27.64 & 23.32 & 16.12 & \textbf{11.30}  \\
 &  
 & IDR & 
 67\% & 31\% & 39\% & 48\% & \textbf{25\%} \\

\cmidrule(lr){2-8}
% RocketChip -- multiply
 & \multirow{2}{*}{multiply}
 & TDT & 
 3.57$^\dagger$ & 17.41 & 12.30 & 21.05 & \textbf{9.71} \\
 &  
 & IDR & 
 71\% & \textbf{30\%} & 33\% & 61\% & 34\% \\

\cmidrule(lr){2-8}
% RocketChip -- spmv
 & \multirow{2}{*}{spmv}
 & TDT & 
 6.02$^\dagger$ & 22.01 & 15.20 & 9.34 & \textbf{8.11} \\
 &  
 & IDR & 
 67\% & 52\% & 31\% & 38\% & \textbf{28\%} \\

\cmidrule(lr){2-8}
% RocketChip -- mm
 & \multirow{2}{*}{mm}
 & TDT & 
 9.31$^\dagger$ & 21.73 & 24.98 & 16.75 & \textbf{9.18} \\
 &  
 & IDR & 
  60\% & 51\% & 48\% & 54\% & \textbf{23\%}  \\

\cmidrule(lr){1-8}

%=========================================================
% BOOM -- coremark
\multirow{18}{*}{\textbf{BOOM}} 
 & \multirow{2}{*}{coremark} 
 & TDT & 
 9.56$^\dagger$ & 32.11 & 27.48 & 35.00 & \textbf{24.14} \\
 &  
 & IDR & 
54\% & 63\% & 42\% & 47\% & \textbf{32\%} \\

\cmidrule(lr){2-8}
% BOOM -- dhrystone
 & \multirow{2}{*}{dhrystone}
 & TDT & 
 11.90$^\dagger$ & 35.00 & 24.92 & 22.18 & \textbf{19.63} \\
 &  
 & IDR & 
 64\% & 57\% & 38\% & 43\% & \textbf{21\%} \\

\cmidrule(lr){2-8}
% BOOM -- rsort
 & \multirow{2}{*}{rsort}
 & TDT & 
 7.56$^\dagger$ & 35.00 & 29.90 & 26.10 & \textbf{14.71} \\
 &  
 & IDR & 
 71\% & 63\% & 45\% & 50\% & \textbf{25\%} \\

\cmidrule(lr){2-8}
% BOOM -- qsort
 & \multirow{2}{*}{qsort}
 & TDT & 
 6.17$^\dagger$ & 32.13 & 22.13 & 28.12 & \textbf{18.71} \\
 &  
 & IDR & 
 71\% & 72\% & 52\% & 36\% & \textbf{15\%} \\

\cmidrule(lr){2-8}
% BOOM -- multiply
 & \multirow{2}{*}{multiply}
 & TDT & 
 7.21$^\dagger$ & 31.61 & 35.00 & \textbf{16.43} & 18.23 \\
 &  
 & IDR & 
 83\% & 52\% & 33\% & \textbf{10\%} & 17\% \\

\cmidrule(lr){2-8}
% BOOM -- spmv
 & \multirow{2}{*}{spmv}
 & TDT & 
 4.42$^\dagger$ & 28.15 & 35.00 & 27.96 & \textbf{9.12} \\
 &  
 & IDR & 
 83\% & 58\% & 39\% & \textbf{36\%} & \textbf{36\%} \\

\cmidrule(lr){2-8}
% BOOM -- mm
 & \multirow{2}{*}{mm}
 & TDT & 
 9.31$^\dagger$ & 21.73 & 24.98 & 16.75 & \textbf{10.18} \\
 &  
 & IDR & 
 60\% & 51\% & 48\% & 54\% & \textbf{23.5\%} \\

%=========================================================

\bottomrule
\end{tabular}
} % end of scalebox
\\
$^\dagger$TDT for HC is not compared with BO methods due to poor design quality
\end{table}

\begin{table}[htbp]
\centering
\footnotesize
\renewcommand{\arraystretch}{0.8}
\caption{Design Quality Measured by Estimated Execution Time (EET) in Milliseconds}
\label{Tab:Experiment_Results_Processor_Performance}
\setlength{\tabcolsep}{4.5pt}
\scalebox{0.8}{
\begin{tabular}{@{}cc cccccc@{}}
\toprule
\textbf{Processor} & 
\textbf{Task} & 
\textbf{HC} & 
\textbf{VBO} & 
\textbf{BE} & 
\textbf{RCBO} & 
\textbf{ASPO} &
\textbf{Default} \\
\midrule

%=========================================================
% EL2 VeeR -- coremark
\multirow{8}{*}{\textbf{EL2 VeeR}} 
 & coremark
 & 9.17 & 9.09 & 9.62 & 9.39 & \textbf{9.18} & 10.17 \\
 % \cmidrule(lr){2-9}

% EL2 VeeR -- dhrystone
 & dhrystone
 & 8.76 & 8.68 & 8.48 & 8.35 & \textbf{7.91} & 9.06 \\
 % \cmidrule(lr){2-9}

% EL2 VeeR -- rsort
 & rsort
 & 6.48 & 5.98 & 6.35 & 5.96 & \textbf{5.82} & 6.42\\

% EL2 VeeR -- qsort
 & qsort
 &
 5.34 & 5.14 & 5.23 & \textbf{5.05} & 5.16 & 5.35 \\

% EL2 VeeR -- multiply
 & multiply
 &
 1.80 & 1.68 & 1.56 & 1.52 & \textbf{1.35} & 1.83 \\

 % EL2 VeeR -- spmv
 & spmv
 &
 1.55 & 1.52 & 1.47 & 1.42 & \textbf{1.35} & 1.57 \\

  % EL2 VeeR -- mm
 & mm
 &
 1.31 & 1.23 & 1.29 & \textbf{1.15} & 1.16 & 1.31 \\

\cmidrule(lr){1-8}

%=========================================================
% RocketChip -- coremark
\multirow{8}{*}{\textbf{RocketChip}} 
 & coremark
 & 8.09 & 8.01 & 7.61 & 7.61 & \textbf{7.02} & 8.09 \\
 % \cmidrule(lr){2-9}

% RocketChip -- dhrystone
 & dhrystone
 & 7.96 & 7.68 & 7.48 & 7.45 & \textbf{7.41} & 7.96 \\
 % \cmidrule(lr){2-9}

% RocketChip -- rsort
 & rsort
 & 5.60 & 5.34 & 5.29 & 5.50 & \textbf{5.09} & 5.61\\

% RocketChip -- qsort
 & qsort
 &
 4.35 & 4.43 & 4.33 & 4.12 & \textbf{4.05} & 4.35 \\

% RocketChip -- multiply
 & multiply
 &
 1.33 & 1.30 & 1.26 & \textbf{1.05} & 1.08 & 1.33 \\

 % RocketChip -- spmv
 & spmv
 &
 1.17 & 1.02 & 0.97 & 1.02 & \textbf{0.93} & 1.17 \\

  % RocketChip -- mm
 & mm
 &
 0.91 & 0.88 & 0.84 & 0.88 & \textbf{0.82}& 0.91 \\

\cmidrule(lr){1-8}
\multirow{8}{*}{\textbf{BOOM}} 
 & coremark
 & 7.17 & 6.89 & 6.62 & 6.79 & \textbf{6.28} & 7.20 \\
 % \cmidrule(lr){2-9}

% BOOM -- dhrystone
 & dhrystone
 & 7.06 & 6.68 & 6.48 & 6.45 & \textbf{6.11} & 7.06 \\
 % \cmidrule(lr){2-9}

% BOOM -- rsort
 & rsort
 & 3.55 & 3.45 & 3.45 & 3.46 & \textbf{3.42} & 3.56\\

% BOOM -- qsort
 & qsort
 &
 3.35 & 3.30 & \textbf{3.12} & 3.27 & \textbf{3.12} & 3.35 \\

% BOOM -- multiply
 & multiply
 & 1.07 & 0.86 & 0.79 & 0.74 & \textbf{0.70} & 1.07 \\

 % BOOM -- spmv
 & spmv
 & 0.87 & 0.78 & 0.80 & \textbf{0.73} & 0.74 & 0.89 \\

  % BOOM -- mm
 & mm
 & 0.73 & 0.75 & 0.66 & 0.72 & \textbf{0.63} & 0.75 \\

\bottomrule
\end{tabular}
} % end of scalebox
\end{table}

% The experimental results, presented in Table~\ref{Tab:Experiment_Results_in_Design_Efficiency} and Table~\ref{Tab:Experiment_Results_Processor_Performance}, compare the design productivity and quality of ASPO against other processor design methods.

Table~\ref{Tab:Experiment_Results_in_Design_Efficiency} highlights the productivity of the processor design methods. Except for the HC method, which converges quickly but does not optimize the processor, ASPO achieves the best performance with the smallest Total Design Time (TDT) and lowest Invalid Design Rate (IDR) in most tasks. For the EL2 VeeR processor with fewer parameter constraints, ASPO achieves a comparable IDR to other methods while still attaining optimal Total Design Time (TDT) thanks to its accelerated evaluation framework. For the more complex RocketChip and BOOM processors with multiple constraints, ASPO significantly lowers IDR, showing its strength in generating valid configurations. As evaluation time increases, ASPO’s TDT advantage becomes more pronounced. For BOOM's optimization on certain benchmarks like ``spmv'', it saves 67.4\% TDT compared with RCBO, 74\% with BE and 67.6\% with VBO. 

Table~\ref{Tab:Experiment_Results_Processor_Performance} focuses on design quality. Processors optimized using ASPO outperform those designed by other methods in 17 out of 21 tasks. In the remaining tasks, ASPO achieves performance comparable to the best alternative methods, demonstrating that it consistently identifies high-performance processor designs. Results show that with ASPO, the optimized BOOM design reduces execution time by 34.6\% on the ``multiply'' benchmark.

\section{Conclusions and Future Work}

This paper presents ASPO, an automated design platform for efficiently configuring soft processors while adhering to FPGA constraints. At the core of ASPO is a Bayesian Optimization (BO) method that incorporates parameter constraints, effectively handles categorical parameters and is aware of evaluation time cost. Additionally, ASPO includes an accelerated evaluation framework with checkpoint data and dynamic checkpoint retrieval for incremental synthesis, significantly reducing the evaluation time. Experimental results demonstrate that ASPO consistently outperforms existing methods, achieving shorter optimization times and higher-quality designs.

Future work includes extending ASPO to support additional architectures, such as System-on-Chip designs and custom instruction processors, for greater application-specific optimization. Additionally, we aim to integrate multi-objective optimization to balance performance, power, and area metrics, further enhancing the platform's versatility.

\section{Acknowledgment}

The support of the United Kingdom EPSRC (grant number UKRI256, EP/V028251/1, EP/N031768/1, EP/S030069/1, and EP/X036006/1), Intel, and AMD is gratefully acknowledged.
% \clearpage
\bibliographystyle{IEEEtran}
\bibliography{Reference_Set}

\end{document}